\documentclass[preprint,12pt]{elsarticle}

\usepackage{amssymb}
\usepackage{pdfpages}
\usepackage{fontawesome}
\usepackage{amsmath}
\usepackage{url} 
\usepackage{multirow}
\usepackage[colorlinks=true, linkcolor=blue, citecolor=blue, urlcolor=blue]{hyperref}

\journal{Information Systems}

\begin{document}

\begin{frontmatter}



\title{Enhancing Urban Data Exploration: Layer Toggling and Visibility-Preserving Lenses for Multi-Attribute Spatial Analysis}


\author[label1,label2,label3,label4]{Karelia Salinas,Luis Gustavo Nonato,Jean-Daniel Fekete,Fernanda Bartolo dos Santos Saran} 

\affiliation[label1,label2,label4]{%
  organization={Institute of Mathematics and Computer Sciences, University of São Paulo (ICMC-USP)},
  addressline={Avenida Trabalhador São-carlense, 400}, 
  city={São Carlos},
  postcode={13566-590}, 
  state={São Paulo},
  country={Brazil}
}
\affiliation[label3]{%
  organization={Université Paris-Saclay, CNRS, Inria, LISN},
  addressline={Bâtiment 660, Rue Noetzlin}, 
  city={Gif-sur-Yvette},
  postcode={91190}, 
  state={Île-de-France},
  country={France}
}

\begin{abstract}
We propose two novel interaction techniques for visualization-assisted exploration of urban data: Layer Toggling and Visibility-Preserving Lenses.
Layer Toggling mitigates visual overload by organizing information into distinct layers while still enabling multi-layer comparisons through controlled overlays. This technique supports focused analysis without sacrificing spatial context and allows users to quickly switch between layers using a dedicated physical button interface.
Visibility-Preserving Lenses, on the other hand, dynamically adapt their size and transparency, enabling users to effectively examine dense spatial regions and temporal attributes in detail. These techniques are meant to support urban data exploration and to improve prediction.

Exploring urban data is essential for understanding complex phenomena related to crime, mobility, and residents' behavior. Equally important is the ability to predict and explain how these phenomena evolve over time, supporting informed urban planning and policymaking. However, navigating urban data in all its complexity is challenging, often resulting in cognitive overload, loss of spatial context, and excessive visual clutter due to the many layers that need to be examined simultaneously. Although layered visualizations aim to mitigate these challenges, they still face limitations with occlusion and effortless comparison across data layers. Additionally, interaction methods are typically confined to mouse-based controls, limiting the fluidity of dynamic exploration.

We validate our visualization tool through a comprehensive user study that measures user performance, cognitive load, and interaction efficiency across multiple devices. Using real-world data from São Paulo, including mobility patterns, climate conditions, and crime statistics, we demonstrate how our approach enhances both exploratory and analytical tasks. The results also show how users perform when playing with different interactive devices, providing guidelines for future developments and improvements.
\end{abstract}

\begin{graphicalabstract}
\begin{figure}[!ht]
   \includegraphics[width=\linewidth]{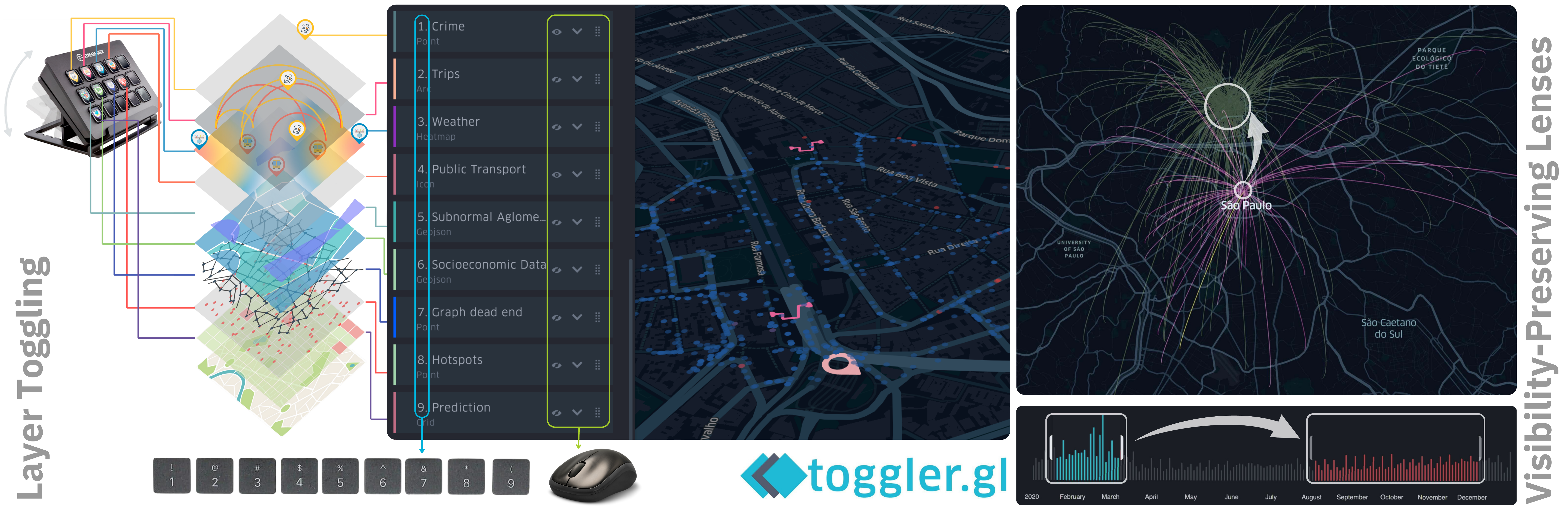}
   \caption{
    Multisource urban databases are rendered as spatially aligned layers that are overlaid to create a comprehensive view. Each layer visualizes a specific type of spatial data using a dense representation. On the left, each of the nine layers is associated with a button in the button box, enabling users to interactively toggle its visibility on the screen and explore spatial correlations between layers, leveraging retinal persistence. Access can also be achieved through keyboard shortcuts corresponding to each layer number or via mouse interactions. 
    In the top-right, the Visibility-Preserving Lens dynamically adjusts the brush radius based on the graph link density at the mouse position, thereby enhancing graph legibility and serving as a spatial filtering mechanism. For temporal data, the size of the range sliders adapts dynamically to the local distribution. Animation can be employed to sweep across the entire range while maintaining controlled legibility. The layers interact and overlap, facilitating the analysis of relationships between various types of data.
   }
   \label{fig:teaser}
\end{figure}
\end{graphicalabstract}

\begin{highlights}
\item Introduces a novel approach to toggle data layers for spatial crime analysis.

\item Proposes visibility-preserving lenses to compare urban attributes with less occlusion.

\item Supports multi-attribute exploration in dense urban visualizations.

\item User study shows improved accuracy and interaction efficiency with proposed tools.

\item Enables better decision-making by enhancing insight into urban spatial patterns.
\end{highlights}


\end{frontmatter}



\section{Introduction}
\label{secIntroduction}
Urban data visualization plays a crucial role in understanding and predicting various phenomena, including accidents, crime, and mobility patterns. As cities become increasingly complex, the ability to visualize multivariate and multimodal data is essential for informed decision-making and effective resource allocation. In the city of São Paulo---Brazil, for instance, the dynamics of taxi trips can intersect with criminal activities involving passengers, drivers, and external parties. To anticipate such behaviors, numerous predictive models have been developed~\cite{garcia2019crimanalyzer,garcia2021cripav,ferreira2013visual,freire2014riding,bogomolov2014once,tekin2021crime}. These models can be evaluated along three critical dimensions: the success rate (How Much?), the underlying reasons for their success (Why?), and the specific locations where they excel (Where?). While machine learning techniques address the ``How Much'' aspect through accuracy metrics, explainability methods tackle the ``Why'' question.
To effectively address the ``Where'' component, we propose leveraging 2D spatial data visualization that integrates multiple attributes and datasets.

Recent advancements in predictive modeling have shifted focus from solely relying on historical data, such as criminal incidents, to incorporating a broader range of urban information, including socioeconomic factors, points of interest, and climate data. This integration gives rise to a comprehensive amount of data serving as feature vectors for machine learning models. From a visualization perspective, this can be represented as a series of overlapping layers, each corresponding to a specific data type (e.g., origin-destination graphs for traffic, and heatmaps for crime and real estate values).

Approaches to visualizing this data often oscillate between two extremes: space multiplexing and the utilization of the entire screen space. Many visual analytic applications traditionally adopt space multiplexing, dividing the screen into multiple sections~\cite{gao2014multi, garcia2021cripav, salinas2022cityhub} to display various maps simultaneously. While this methodology can enable the visualization of diverse datasets, it limits the overall analysis of the attributes, making it difficult, for instance, to observe spatial correlations. Conversely, the use of the entire screen space allows for a more panoramic information view, facilitating the perception of both local and global correlations among features. However, this method can lead to visual clutter, complicating effective analysis due to the accumulation of data within the same spatial framework. Solutions such as data aggregation~\cite{von2015mobilitygraphs} and filtering~\cite{guo2009flow} techniques have been proposed to mitigate these challenges. Nonetheless, data aggregation may smooth out or obscure important patterns, while static filtering can risk concealing broader contexts.

In this work, we propose a novel methodology where each data layer occupies the entire screen space, thereby avoiding spatial multiplexing, maintaining high spatial resolution and correlation between layers. We efficiently manage layer visibility and order using physical buttons or keys, and utilizing dynamic lenses for filtering to ensure that the layers remain readable. Our approach incorporates \emph{layer toggling}, where each layer is assigned to a key on an auxiliary device, allowing users to maintain their view focus while easily comparing features. Additionally, we introduce a dynamic filter for temporal and spatial data, featuring a lens that adapts to data density and a window that adjusts based on the distribution of temporal data.
Developing a novel method for visual analysis is essential in this context, particularly in light of the prevalent tendency to divide screens into multiple views, which can hinder the visualization of the whole data and miss correlations between information layers. Moreover, when the output of predictive models is part of the analysis, identifying areas with poor predictions is vital for analyzing the causes and improving model performance. 

To validate our methodology, we utilize a dataset focused on the city of São Paulo (denoted São Paulo for now on), the largest city in South America, which offers a rich diversity of open data. This extensive dataset enables the incorporation of varied information into our analytical framework.

Our methodology consists of several key steps. We begin by aggregating data using the CityHub library~\cite{salinas2022cityhub}. Next, we apply a machine-learning classification algorithm to predict whether a taxi trip will involve a crime occurrence. 
For visual presentation of the datasets and prediction results, we utilize the \texttt{kepler.gl} software~\cite{he2018beautiful}, incorporating customized filtering and navigation functions, along with a ``stream deck'' button box for efficient layer toggling. Finally, we validate the effectiveness of our approach through an in-depth analysis of São Paulo, which yields valuable insights.

We tested our approach with users divided into several groups to evaluate whether the use of layers aids urban analysis and to assess differences in performance and efficiency with three different auxiliary devices for layer toggling. In a second phase, we compared static versus dynamic filters. 

This user study demonstrated that the use of layered data visualization significantly enhances exploration and understanding; however, no significant differences were found between the various devices used. Additionally, it was confirmed that dynamic filters facilitated tasks compared to static filters in terms of execution time.

In summary, our contributions include:

\begin{itemize}
\item Proposing a novel analysis method that organizes data into layers.
\item Introducing the Layer Toggling technique for navigation.
\item Implementing density-based dynamic filtering for spatial and temporal data.
\item Providing a visual evaluation of predictive system performance, highlighting areas of success and failure. 
\item Conducting user tests to assess the effectiveness of layered visualization in facilitating exploration and evaluating the impact of different devices on layer toggling, as well as comparing the efficiency of static and dynamic filters using statistical analysis. 
\end{itemize}

\section{Related Work}
\label{secRelatedWork}
Urban data visualization often involves complex datasets that can lead to visual clutter, obscuring meaningful insights. To effectively address this issue, it is essential to explore various taxonomies of urban data representation. Boyandin et al.~\cite{boyandin2013visualization} and Gu et al.~\cite{gu2023classification} provide a comprehensive classification of data representation approaches, which can be broadly categorized into cartographic contexts and other diagrammatic forms.
A widely investigated type of urban data is origin-destination, where several representations and visualizations stand out. Flow Maps, for instance, successfully encode multiple data components using flow symbols, allowing for the efficient visualization of both directed and undirected movement~\cite{dent1999cartography}. In contrast, Thematic Maps do not employ line symbols to indicate flow; instead, they utilize alternative visual variables to represent the flow between origins and destinations~\cite{bertin1967semiology}. Other diagrammatic approaches include Matrices, where columns and rows correspond to origins and destinations, with entries indicating flows~\cite{gu2023classification}. Node-Link Diagrams use node symbols for origins and destinations, while linear symbols represent flows, employing various layouts such as Arc Diagrams~\cite{wattenberg2002arc}, Alluvial Diagrams, Circos~\cite{krzywinski2009circos}, and Hive Plots~\cite{krzywinski2012hive}.

Despite the advances, the cluttering problem remains a significant challenge in urban data visualization~\cite{guo2014origin}. This issue encompasses several facets, including the cluttering problem itself, the Modifiable Area Unit Problem (MAUP), where different aggregations may reveal distinct patterns, and the Normalization (or Size-Difference) Problem, which refers to the challenge of accurately representing data when the sizes of visual elements do not correspond proportionally to the underlying values they represent. This discrepancy can lead to misinterpretations and skewed understandings of the data, making it difficult for users to draw accurate conclusions from the visualizations.
Various techniques have emerged to combat clutter, such as filtering Origin-Destination (OD) data and optimizing symbols and layout through location aggregation, which utilizes spatial clustering~\cite{adrienko2010spatial}, graph partitioning~\cite{guo2009flow}, and discretization in high-level administrative units. Additional methods include layout adjustment techniques~\cite{buchin2011flow}, edge bundling algorithms~\cite{cui2008geometry}, and interactive triangular irregular networks (TIN) modification~\cite{dakowicz2005interactive}. However, these techniques can obscure significant data, smooth patterns, and lead to misinterpretations. Maintaining data integrity without altering the underlying information is crucial, but can lead to visual overload, which we propose to manage through the use of layers and the application of dynamic filters.
In our previous work~\cite{salinas2024navigating}, we introduced a preliminary implementation of layer toggling and visibility-preserving lenses. The present work extends the discussion by providing a detailed explanation of the data sources, preprocessing methods, and the structural modeling process, along with implementation details. Additionally, we have improved the visualizations by enhancing tooltips, optimizing color contrast, refining user feedback within the button box, and making the distinction between activated and deactivated buttons more noticeable. Furthermore, in our prior study, we conducted an experiment with ten users who tested three different devices for layer toggling and two filtering approaches, dynamic and static, ultimately favoring the button box and the dynamic filter. In this work, we expand the number of participants, segment users into different groups, quantitatively measure their performance (including task completion time and click count), and refine the result analysis phase to better justify design decisions.
In this work, we address the cluttering and spatial correlation problems with two interactive techniques: layer toggling and visibility-preserving lenses.

\subsection{Layer Navigation}
Urban data visualization has seen significant advancements in recent years. However, many existing methodologies reveal critical limitations that adversely impact user experience. Techniques for layered navigation often fragment the display into multiple modules, resulting in cognitive overload and impeding a comprehensive understanding of the presented data. Kraak and Ormeling~\cite{kraak2010cartography}, Gao et al.~\cite{gao2014multi}, García et al.~\cite{garcia2021cripav}, and Salinas et al.~\cite{salinas2022cityhub} collectively emphasize that subdividing the visual space restricts the overall analysis of attributes or datasets, complicating the identification of spatial correlations.

To address integration challenges, some works focus on unifying heterogeneous urban data sources. Wang et al.\cite{wang2024urbandatalayer} propose the UrbanDataLayer, which fuses urban data with spatio-temporal base layers through a unified pipeline, improving dataset cohesion but still facing issues of visual clarity at scale. Similarly, Sideris et al.\cite{sideris2024enhancing} employ graph-based convolutional neural networks to model complex relationships in urban data, advancing analytical capabilities at the cost of computational complexity.

Other studies prioritize reducing visual clutter and improving user navigation. Bentlin~\cite{bentlin2024urbanlayer} advocates for graphical reduction and selective removal of unnecessary layers to highlight meaningful relationships, though this risks oversimplifying data and omitting critical details. Zhou and Hsu~\cite{zhou2018layered} introduce interactive layering techniques to enhance engagement, but these may inadvertently increase fragmentation and distract from core analysis tasks.

The complexity of visualizing three-dimensional urban data adds further challenges. Miranda et al.\cite{miranda2024state} discuss the importance of direct associations between data layers in 3D environments to boost user understanding, but warn of potential cognitive overload inherent to such representations. Wagner et al.\cite{wagner2024reimagining} explore immersive analytics via a space-time cube metaphor, enhancing spatial comprehension yet complicating navigation due to its immersive design.

Finally, foundational perspectives by Dodge and Kitchin~\cite{dodge2013outlines} highlight visualization’s essential role in geographic information science but do not offer concrete solutions for seamless multi-layer navigation.

Leveraging the entire screen for a single visualization enhances the clarity of both local and global relationships among features, although it may introduce visual clutter and challenges related to data aggregation.

To address these issues, we implement a visualization system that facilitates \textit{eye fixing}, eliminating the need to subdivide the display into multiple modules. We propose a unified screen with layered visualizations that provide a more coherent and less distracting navigation experience. Furthermore, we incorporate an \textit{external button box} equipped with intuitive controls for layer manipulation, allowing users to interact with the data seamlessly without losing eye focus on their primary objectives. This innovative design significantly enhances user engagement with relevant information while reducing the cognitive burden typically associated with split displays.

\subsection{Lenses}

Lenses, as discussed by Tominski et al.~\cite{tominski2014survey}, provide dynamic filtering capabilities that allow users to focus on relevant data while minimizing clutter. 
The application of lenses as a filtering mechanism in data visualization is prevalent; however, many contemporary approaches exhibit deficiencies in adaptability and usability. Traditional filtering methods frequently fail to dynamically adjust to user needs, leading to information saturation. Bach and Carpendale~\cite{bach2016role} explore the potential of lenses to enhance data exploration, yet they do not sufficiently address how the lack of adaptability can detrimentally affect user experience. Similarly, Keim and Ward~\cite{keim2009visual} highlight various filtering applications but overlook the necessity of maintaining a balanced data representation on-screen to avoid overwhelming users.

Integrating lenses into urban data visualization can significantly enhance user experience by enabling dynamic filtering tailored to individual needs. Fischer and Hegarty~\cite{fischer2015temporal} investigate the use of temporal lenses, contributing valuable insights into time-dependent data visualization. However, a comprehensive solution that adapts to the evolving requirements of users remains elusive.

In this work, we implement adaptive filters that adjust to user preferences while ensuring a balanced presentation of data, thus preventing information overload. This approach not only enhances usability but also optimizes the comprehension of the visualized information.
By merging advanced layered navigation techniques with adaptive lens functionality, this work represents a substantial advancement in urban data visualization, addressing the shortcomings of existing methodologies and incorporating user-centered design principles. This research improves data interpretation and fosters a more effective and engaging user experience.

\section{Methodology}
\label{sec:approach}
\subsection{Design Requirements and Tasks}
\label{sec:requirements}
The development of our tool is rooted in extensive experience working with various entities in the management of urban data, particularly in the manipulation of origin-destination datasets. Through this work, we identified several pressing needs when presenting information to transportation analysts who may lack the expertise to navigate complex software systems. To address these challenges, our tool and its enhancements are guided by the following key requirements:

\begin{itemize}
    \item \textbf{R1.} Maintain resolution by avoiding the fragmentation of the screen into modules, ensuring that all information is displayed in a single, cohesive visualization.
    \item \textbf{R2.} Utilize cartographic maps for data representation, as this approach is more intuitive for transportation professionals and facilitates better understanding of spatial relationships.
    \item \textbf{R3.} Enhance user interaction with the tool to support eye fixation, thereby minimizing distractions from keyboard manipulation and manual movements.
    \item \textbf{R4.} Incorporate necessary filters for the analysis of spatio-temporal data, allowing users to delve deeper into the insights provided by the data.
    \item \textbf{R5.} Spatially represent the results of predictive models to identify failure zones, uncover correlations between layers, and highlight the significance of features that contribute to the predictive outcomes.
\end{itemize}

Our tool is designed to fulfill two primary tasks, each tailored to specific user types, as illustrated in \autoref{fig:table}.

\begin{itemize}
    \item \textbf{T1. Exploration:} This task focuses on inexperienced users, such as regular citizen, who may wish to explore the conditions of specific neighborhoods to assess their suitability for living. This functionality enables users to gain valuable insights into various urban factors that impact their quality of life.
    \item \textbf{T2. Analysis:} This task is aimed at analyst users, including data analysts and specialists in urbanism, transportation, and law enforcement. These users examine the relationships among diverse datasets to identify patterns of abnormal behavior, such as crime and traffic issues. Additionally, the tool serves as a critical resource for evaluating the effectiveness of predictive models.
\end{itemize}

In summary, our methodology not only addresses the technical requirements for effective data visualization but also considers the diverse needs of users, ensuring that both regular users and analysts can derive meaningful insights from urban data. \autoref{tab:layer_relationships} presents which requirements and tasks must be observed when handling data layers and the different functionalities to be implemented to attend the requirements and tasks.

\begin{table}[t]
    \centering
    \resizebox{\columnwidth}{!}{ 
        \begin{tabular}{|l|c|c|c|c|c|c|c|}
            \hline
            \textbf{Layer / Resource} & \textbf{R1} & \textbf{R2} & \textbf{R3} & \textbf{R4} & \textbf{R5} & \textbf{T1} & \textbf{T2} \\
            \hline
            Crime Layer & \checkmark & \checkmark &  &  &  & \checkmark &  \\
            \hline
            Taxi Trips Layer & \checkmark & \checkmark &  &  &  & \checkmark &  \\
            \hline
            Weather Layer & \checkmark & \checkmark &  &  &  & \checkmark &  \\
            \hline
            Public Transportation Layer & \checkmark & \checkmark &  &  &  & \checkmark &  \\
            \hline
            Favelas Layer & \checkmark & \checkmark &  &  &  & \checkmark &  \\
            \hline
            Socioeconomic Layer & \checkmark & \checkmark &  &  &  & \checkmark &  \\
            \hline
            Graph Layer & \checkmark & \checkmark &  &  &  &  & \checkmark  \\
            \hline
            Hotspots Layer & \checkmark & \checkmark &  &  &  &  & \checkmark \\
            \hline
            Prediction Layer & \checkmark & \checkmark &  &  & \checkmark &  & \checkmark \\
            \hline
            Layer Toggling & \checkmark & \checkmark & \checkmark &  &  & \checkmark & \checkmark \\
            \hline
            Visibility-Preserving Lenses &  &  &  & \checkmark &  & \checkmark & \checkmark \\
            \hline
            Correlation Matrix &  &  &  &  & \checkmark &  & \checkmark \\
            \hline
            Shapley Values &  &  &  &  & \checkmark &  & \checkmark \\
            \hline
        \end{tabular}
    }
    \caption{Relationship between layers and resources with the requirements (R) they fulfill and the tasks (T) they address. See more details in \autoref{sec:requirements}.}
    \label{tab:layer_relationships}
\end{table}

\section{Kepler.gl Tool} 
\label{sec:Kepler}

To create this visualization, we built upon the Kepler.gl project~\cite{kepler}. Kepler is an advanced data visualization and analysis platform designed to enable the exploration of large-scale spatial and temporal datasets. It emphasizes interactive tools that empower users to intuitively identify patterns and relationships within their data~\cite{kepler}.

Kepler is constructed using modern web technologies, incorporating frameworks such as React for the user interface and deck.gl, a data visualization library developed by Uber's Data Visualization team. It also leverages WebGL and react-mapbox-gl for rendering dynamic, high-performance visualizations. This robust architecture ensures that Kepler can handle complex datasets with efficiency and responsiveness. For our implementation, we used Kepler.gl v2.5.0, an actively maintained version that continues to receive updates and enhancements from its creators. We customized several of its modules, taking full advantage of the tool's open-source nature.

Key features of Kepler include:

\begin{itemize}
\item \textbf{Spatial Data Visualization:} Kepler enables the display of geospatial data on interactive maps, making it easy to identify trends, anomalies, and geographic patterns. 
\item \textbf{Interactivity:} The platform allows users to manipulate visualizations in real-time, adjusting parameters and filters to explore the data from multiple perspectives. 
\item \textbf{Integration of Multiple Data Sources:} Kepler supports the integration of diverse datasets, facilitating more comprehensive and contextualized analyses. 
\item \textbf{Ease of Use:} Designed for accessibility, Kepler's user interface is intuitive enough for both technical and non-technical users. 
\item \textbf{Modular Architecture:} Kepler's modular design makes it easy to extend and integrate new features. The backend is powered by Node.js, ensuring scalability and high performance when handling large datasets. 
\item \textbf{Open Source:} As an open-source project, Kepler encourages contributions from developers and allows for customization to suit specific needs. 
\end{itemize}

This system is particularly valuable in fields such as urban planning, environmental management, and scientific research, where powerful data visualization tools can lead to better decision-making and groundbreaking discoveries.

\subsection{Data}
\label{sec:data}
\begin{figure*}[h!]
\centering
\includegraphics[width=0.95\textwidth]{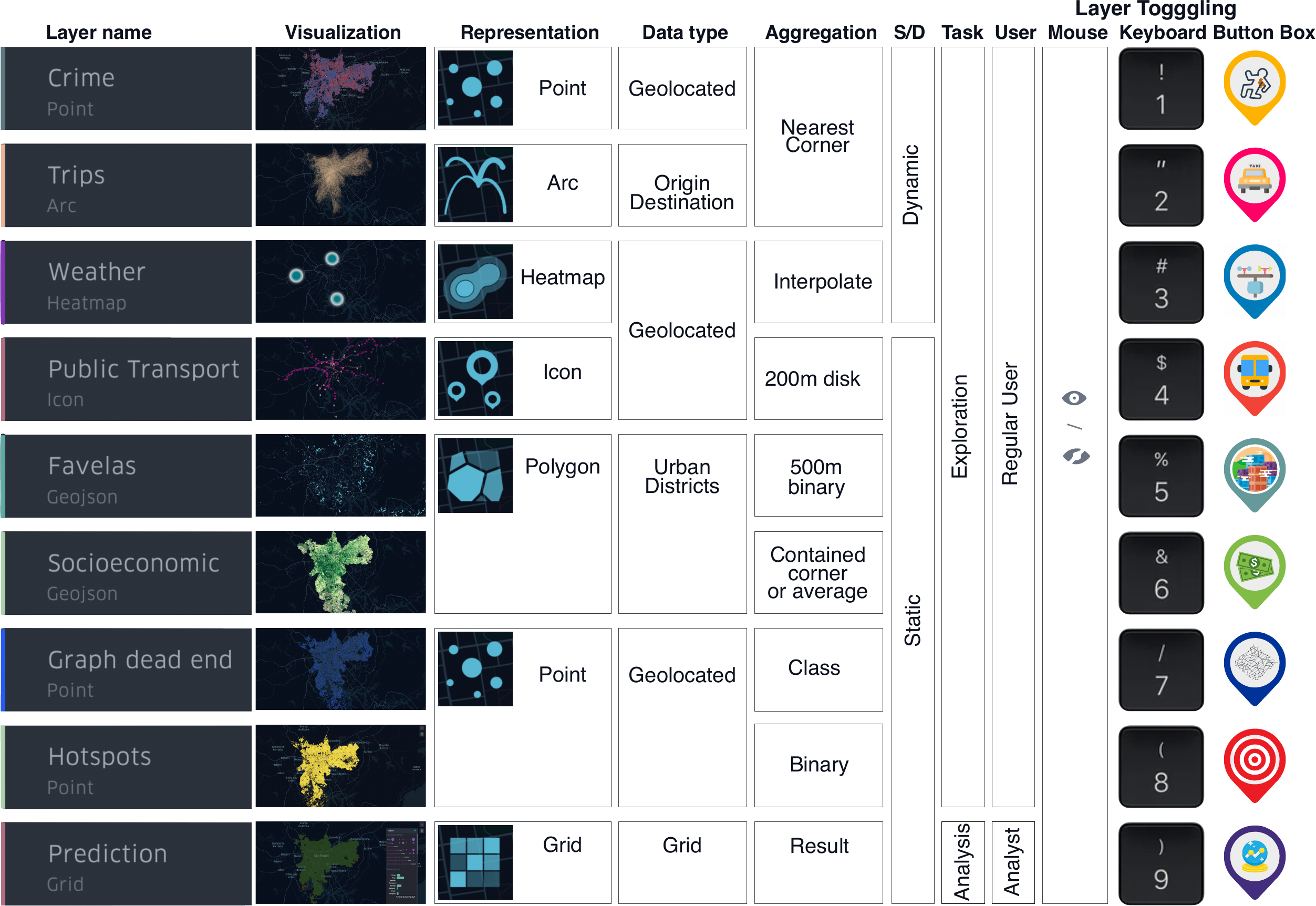}
\caption{%
The layers under consideration, each corresponding to a distinct database. The first column lists the buttons as they appear in the interface, while the second column displays the associated visual representation. The third column illustrates the data representations based on the available types in Kepler, such as points and arcs. The fourth column denotes the data type within each database, including geolocated data and origin-destination.
The fifth column shows how the data is integrated into the spatial domain representation, which is the street a graph in our case. The integration gives rise to feature vectors at each vertex of the graph (street corner) used in the predictive model. The sixth column classifies the data as dynamic or static, depending on update frequency and temporal variability. The seventh and eighth columns outline the tasks associated with each layer and the corresponding potential user. Finally, the last three columns describe how the interaction with different devices for toggling is performed, which can be accomplished using a mouse, keyboard shortcuts, or the button box, with each layer represented by an icon. 
}
\label{fig:table}
\end{figure*}

We collect São Paulo data from various sources and organize them into distinct layers within our analytical tool. Each database is tailored to a specific data type and representation. For example, geolocated data is depicted as points, a feature already integrated into Kepler. 
We categorize the data into two primary types: dynamic and static. Dynamic data, such as crime statistics, taxi trips, and weather conditions, changes frequently over time. \autoref{fig:table} provides a comprehensive summary of the data types, their respective representations across each layer, how they are aggregated, which user is expected to handle them, and how the respective layer can be accessed in the different devices.

In the following, we provide a detailed description of each database, elucidating the source, visual representation, and the methodology employed to integrate the data into the graph nodes for predictive tasks. The graph is constructed based on the city's street map, where nodes represent street corners and edges represent the streets connecting them.

\subsubsection{Crime}
Crime data was obtained from the Open Data Portal of the Secretary of Public Security of the State of São Paulo (\url{www.ssp.sp.gov.br}). Each crime record includes information such as geolocation (latitude and longitude), date and time of occurrence, and the type of crime. For this study, we specifically focused on crimes categorized as vehicle and mobile phone thefts occurring in 2020. In our visualization, crime data is represented as geolocated points, with red points indicating vehicle thefts and blue points representing mobile phone thefts. Furthermore, crime records were assigned to the nodes of the street graph using the nearest corner approach, as proposed by Garcia-Zanabria et al.~\cite{garcia2021cripav}.

\subsubsection{Taxi Trips}
Based on population density in census tracts, we generate a synthetic ride-hailing dataset that emulates the taxi data, which comprises 87,000 records labeled as ``regular'' or ``occurrence''. The former represents trips without any criminal incidents, such as robbery or theft, while the latter corresponds to trips in which such incidents occurred. Normalized population density is used as probabilities to sample randomly census track as origin and destination, using the distance to crime hotspots to label trips as ``occurrence'' or ``regular'' in a ratio of about 1:90 when the distance from the origin or destination to the hotspot is smaller them a 500m. This procedure generates a highly imbalanced dataset.
Each trip record includes information such as the closest street corner (graph node) to the origin and destination computed from the location of the census track, the period of day---morning [6 AM-12 PM], afternoon [12 PM-6 PM], night [6 PM-12 AM], or dawn [12 AM-6 AM]---(more samples are drawn in the working hours), the day of the week (less samples are drawn in the weekends), and the trip month (about the same number of samples in each month). 
We represent the origin-destination data as arcs connecting the starting and ending points of the trips.

The predictive task focuses on this dataset, aiming to forecast whether a trip will involve a criminal occurrence. To achieve this, we construct a feature vector for both the origin and destination, concatenate them, and process the combined vector through a binary response predictive model to determine the likelihood of crime. This layer not only enriches the graph but also leverages it for predictive purposes. Additionally, taxi pick-up and drop-off count can be recorded at each node, following the same nearest corner criterion.

\subsubsection{Weather}
Temperature and rainfall data were collected from three weather stations around the city in 2020, sourced from the Brazilian National Meteorology Institute. This data is spatially sparse but temporally rich. In our map visualization, we pinpoint the exact locations of the three weather stations in São Paulo, located at the district of Barueri, Interlagos, and Mirante. The climate value at each graph node is obtained by linearly interpolating information from the three stations, using as weight coefficients the inverse of the distance between the location and the three stations.

\subsubsection{Public Transportation}
The location and types of public transportation facilities (bus stops, terminals, subway, and train stations) were obtained from the Geosampa portal (\url{geosampa.prefeitura.sp.gov.br}). This data is geolocated, and we utilize distinct icon representations for each category on the map. The count of each type of transportation facility within a 200-meter radius centered on each node is used as the data.

\subsubsection{Favelas - Subnormal Agglomerates}
Geolocation data for subnormal agglomerates was obtained from the IBGE repository (\url{www.ibge.gov.br}) for 2019. We visually represent this data as polygons, utilizing GeoJSON data for these urban districts. A binary variable is assigned to each street graph node, indicating whether it is located within 500 meters of a subnormal agglomerate.

\subsubsection{Socioeconomic Data}
Socioeconomic data were collected from the Brazilian Census database (\url{www.ibge.gov.br}) for 2010. This data is originally aggregated in census tracts represented by polygonal curve. Seven socioeconomic indicators were considered: average household income, average householder income, unemployment rate among householders, literacy rate for children aged 7 to 15 years, percentage of residents under 18 years, percentage of residents aged 18 to 65 years, and percentage of residents over 65 years. To aggregate this data in the graph, each corner captures all data from the census sector to which it belongs; if located on the boundary between multiple sectors, it stores the average of all the data.

\subsubsection{Graph Classification}
In addition to the variables described above, each node of the street graph is classified as \emph{dead end}, \emph{near dead end}, or \emph{regular}. A node is classified as a \emph{dead end} if it is connected to a node of degree one, indicating a street without an exit. If there is a dead-end node within a 100-meter path from a node, that node is classified as a \emph{near dead end}. Otherwise, the node is classified as \emph{regular}. These classifications are represented as points in the visualization. In the graph, each node preserves its value.

\subsubsection{Crime Hotspot Identification}
Hotspot identification techniques aim to detect locations with a high risk of crime. Identifying crime hotspots is a significant task in crime mapping~\cite{chainey2013gis,eck2005mapping,eftelioglu2020crime}. In our analysis, we employ a Markov Model to detect hotspots~\cite{robertson2022predicting,mondal2022crime} 
which are also represented as points in the map. The graph nodes preserve their value.

\subsubsection{Prediction}
Effective classification models depend on robust feature vectors. A critical challenge in our context is the highly imbalanced nature of the ride-hailing dataset. Consequently, we conducted a rigorous experiment to select an appropriate classification model that accommodates this imbalance. Among the classification algorithms considered for imbalanced datasets, XGBoost~\cite{chen2016xgboost} combined with Random Undersampling~\cite{he2009learning} yielded the best results, achieving a G-mean performance of 0.89.

Each database above corresponds to a data layer, from which we generate a feature vector for each node of the street graph. The predictive task relies on origin-destination trips, where we concatenate the feature vector for the origin with that of the destination and input it into the classification model to predict whether a crime occurrence will take place during a taxi trip. This prediction is then compared with the labels in our trip database; correct predictions are counted as positive points, while incorrect predictions are marked as failures. Subsequently, a regular grid is used as domain discretization where the count of  successes and failures are aggregated with each grid cell. The result is visualized in a three-dimensional view, where the height corresponds to the count within each cell, with green color indicating success and red failure. This layer fulfills requirement R5 and supports task T2 related to the analysis.

\autoref{fig:prediction} illustrates the prediction grid, where we observe that the central area exhibits a higher concentration of correct predictions. Conversely, the peripheral regions demonstrate a greater number of failures. The height of the grid is particularly significant, as it correlates with the larger volume of trips occurring in the central area. In contrast, the peripheral areas, where the model tends to underperform, reflect a lower density of trips.

\section{Visualization Design}
\label{sec:visualization}
The proposed solution has been built upon the open-source project Kepler.gl~\cite{kepler} described in \autoref{sec:Kepler}, although the proposed solutions can also be adaptable to other platforms. The implementation adheres to the requirements outlined in \autoref{sec:requirements} and effectively addresses the exploration and analysis tasks. Specifically, the proposed solutions to tackle data layer-based entire screen visualization and adaptive lens filtering problems has been implemented as follows:

\subsection{Layer Toggling}
Implemented to satisfy requirement R1, this feature accommodates the diverse databases, while been able to deal with dynamic and static data, properly handling the different data types and visual representations. We treat each database as a layer, as illustrated in \autoref{fig:teaser}. These layers occupy the entire screen viewport and can overlapped during visualization. To facilitate seamless transitions between layers while maintaining retinal persistence, we developed a technique known as Layer Toggling. This technique allows for instantaneous layer changes without losing spatial focus, delegating the task to manual control rather than visual attention. We implemented three versions of this technique in order to accommodate its operation in  different devices - mouse, keyboard, and button box, as detailed in the following.

\begin{itemize}
\item Mouse Interaction: An icon is added to each layer button, positioned at the rightmost part adjacent to the map. This setup enables the layer to be displayed when user click with mouse in the symbol \faEye\, and turned off when click in \faEyeSlash.

\item Keyboard Interaction: A key on the computer keyboard is assigned to each layer. Users have two options: assigning a number to each layer for easier correspondence or assign customized keys. The later is useful when users have keyboards with tactile reliefs for intuitive navigation, facilitation to navigate layers without diverting their gaze from the screen.

\item Button Box: While the previously mentioned options are suitable for traditional computers, we found that employing an additional device could significantly enhances navigation capabilities. Thus, we selected the ``Elgato Stream Deck Classic''~\cite{streamdeck}. This live production controller features 15 customizable LCD keys and an adjustable stand. Unlike other versions, it allows users to calibrate the tilt angle and is priced under 200 euros.

The Stream Deck Classic is organized as a 5x3 key layout, as depicted on the top-left in \autoref{fig:teaser}. For ergonomic purposes, we utilize only the four horizontal buttons from each of the three columns, each assigned to the index, middle, ring, and little fingers. This configuration allows fingers to navigate across the nine activation buttons. Each button corresponds to a layer, and to aid memory retention, we include an icon representing the content alongside the layer name, which is visible from all tilt angles.
\end{itemize}

Notice that all three interactive resources above aims to address requirement R3, although the use of keyboard with without tactile reliefs diverges a bit from this goal.

\subsection{Visibility-Preserving Lenses}
We have implemented two types of lenses for dynamic filtering, which adjust based on an initial parameter that can be either spatial or temporal. These lenses have been developed to address requirement R4.

\begin{itemize}
\item Spatial Lens: The first type of lens is a brush that allows exploration of a specified number of points, such as one hundred crime incidents. When the cursor is moved over the map, this brush highlights only groups of one hundred points. In areas with higher point concentrations, the brush size decreases, while in areas with fewer points, it increases.

To implement this feature, we drew inspiration from the work~\cite{knnquad}, which combines Quad Tree data structure~\cite{finkel1974quad} with  K-Nearest Neighbor (KNN) clustering 
algorithm~\cite{cover1967nearest}. \autoref{fig:knnquad} illustrates the integration of both methods.

Data points are inserted into a Quad Tree, a process applicable only to layers with geolocated or origin-destination data types, as they possess specific latitude and longitude coordinates on the map.

In the interface, we include a slider to specify the number of points to be kept in the visualization. This, combined with the cursor's position, serves as input for the KNN algorithm, allowing to search for the "k" closest points. We then save the location of the last point found, e.g., the most distant from the cursor—and measure the distance between both points to establish the radius of the brush.

\begin{figure}[tb]
  \centering 
  \includegraphics[width=0.5\columnwidth]{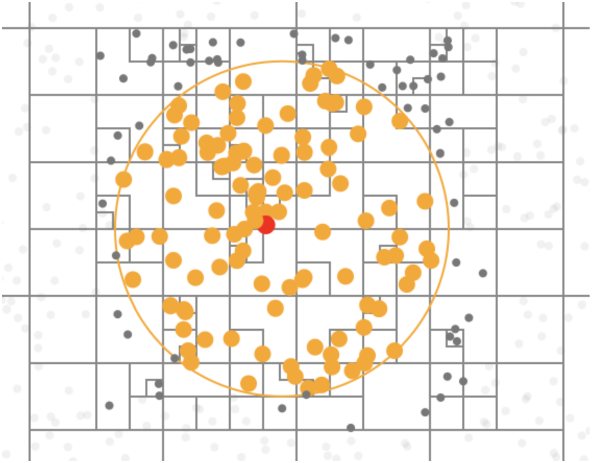}
  \caption{%
  Division of space using a Quad Tree, represented by gray quadrants. The application of KNN to find the ``n'' points closest to the cursor is highlighted in red. Explored points appear in gray, while found points are shown in orange.
  }
  \label{fig:knnquad}
\end{figure}

\item Temporal Lens: Temporal data, such as the distribution of crimes by month, is depicted as a histogram within a folding window. The implemented lens enables users to select a range of bars to display, for instance, crimes in the months of February and March. Additionally, Kepler offers an animation feature that begins with the initial range input (e.g., two months) and progresses to display subsequent months until the entire distribution is covered. Upon completion, the animation seamlessly loops back to the beginning.

We have also incorporated several improvements into this tool. For instance, suppose we want to compare crime rates in February and March to other periods of the year. To achieve this, the dynamic lens accepts inputs in terms of occurrence count or density. Consequently, the animation's progression is no longer fixed at two months but dynamically adjusts to represent time intervals equivalent to the occurrences in February and March. The cumulative sum algorithm~\cite{knuth1997art} facilitates this process. As the animation progresses, each subsequent bar is added until the initially set amount is reached, with the first bar being subtracted. This approach prevents screen overload by consistently displaying the same quantity of data in each frame. Furthermore, to maintain continuity in the visualization, the animation reverses direction upon reaching the end of the histogram, rather than looping back to the beginning.
\end{itemize}

\begin{figure}[tb]
\centering
\includegraphics[width=\linewidth]{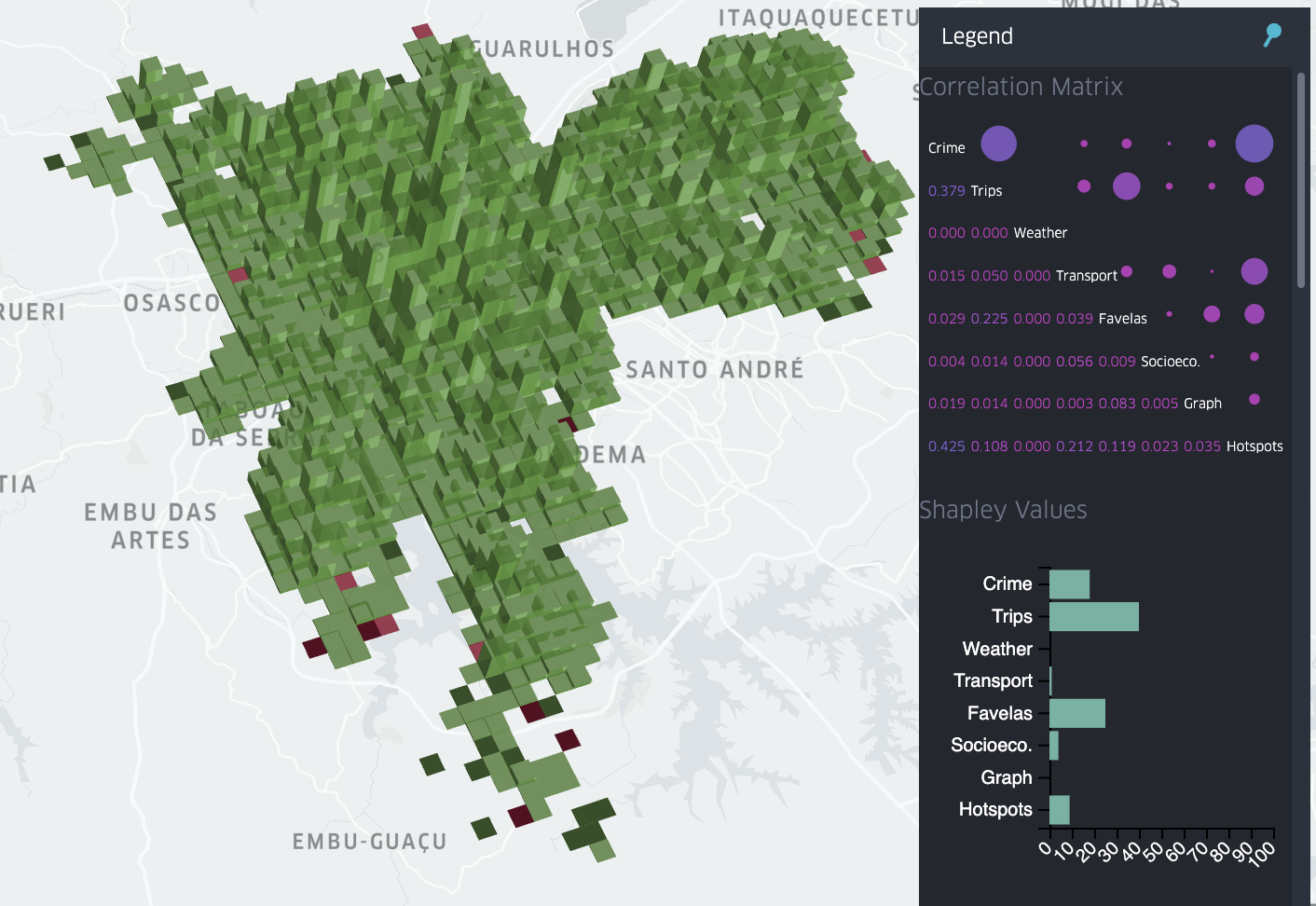}
\caption{%
Resources for Analytical Tasks: The figure illustrates the spatial representation of the classification model's behavior, addressing the initial question of where the model underperforms, specifically in the peripheral areas. Additionally, on the right, other resources for analysts are presented, including the correlation matrix and the Shapley values. 
}
\label{fig:prediction}
\end{figure}

Additionally, we have implemented several other customized tools to improve computational performance, including:

\begin{itemize}
\item Efficient Handling of Large Parquet Files: We integrated a database reader and manager utilizing the DuckDB library~\cite{duckdb}. This integration significantly enhances the speed of reading and querying, addressing challenges associated with big data.

\item Correlation Matrix: To simplify the selection of layers, we incorporated a correlation matrix of the layers into the map legend, as seen in \autoref{fig:prediction}. This enables expert analysts to seamlessly switch between the most correlated layers to elucidate hypothesis, thereby addressing requirement R5. The correlation matrix is constructed by first computing the pairwise correlations (using Pearson’s coefficient) of all 52 features that contribute to the predictor. These features are grouped into eight thematic layers. To synthesize the matrix at the layer level, we compute the average of the correlation values within each layer. The resulting reduced matrix, shown in the legend, summarizes the average intra-layer correlations.

\item Shapley Values: We also included a bar chart displaying Shapley values~\cite{roth1988introduction} 
in the legend to indicate which characteristics contribute most significantly to the classification model decison. This highlights important layers that should be prioritized in the analysis. We consider only the absolute values, as the direction of contribution (positive or negative) is less relevant than the percentage of contribution or importance. This tool also contributes to the fulfillment of requirement R5.
\end{itemize}

\section{Illustrative Case Study}
In this section we present a scenario in which a user utilizes the tool to fulfill both exploration and analysis tasks. Initially, the user engages with the tool as a regular user for exploratory purposes, followed by an analytical approach as a data analyst. This illustrative case study demonstrates how the proposed methodology can support urban data exploration and provide valuable insights for users.

Consider a scenario where an individual is searching for an apartment to rent in São Paulo, specifically around Paulista Avenue. He is interested in assessing the neighborhood's security, transportation facilities, and overall cost of living. He uses our approach to \emph{explore} the city. First, he insert an address in the neighborhood he is interested in, resulting in an icon placed at that location.

To evaluate the security aspect, he activates the ``Crime'' layer and observes a notable concentration of crimes in the area. Knowing that, he decides to go deeper and analyze if crimes are, in fact, concentrated near the location he choose. For that, he uses the spatial lenses, hovering the mouse along the streets and avenues. Particularly nearby his specified address the brush contracts, indicating a greater concentration of crime. This observation provides valuable insights into the distribution and density of crime occurrences along the street. He wants to go even further, seeking know the most problematic hours to avoid them. He uses the temporal filter and see the distribution of crimes by hours, having a bigger concentration at night. He realizes that the quantity of crimes at 8 pm is equivalent to all the dawn [12am-6am] occurrences, that thanks to the temporal dynamic lens. Now he knows which areas and times to avoid to be safer.

Regarding mobility, he investigate public transportation facilities near his address, so he swaps instantaneously to the ``Public transportation'' layer; he doesn't lose the view of the particular address because he has the auxiliary keyboard in his left hand with fingers placed in the layers of his interest. In that layer, he saw many nearby bus stops and three subway stations. Although the place is well served in terms of public transportation, what would happen if he decides to take a taxi? Is it safe in that zone? He switches to the "Taxi trips" layer, where he can observe highlighted all the taxi trips with some criminal occurrences. He raises the hypothesis that the economic level of people living in that zone might be attacking criminals to the area. 

He decides to confirm his hypothesis and toggled to the ``Socioeconomic'' layer. The darker green color of the neighborhood reveals a high income rate, corroborating his suspicions. Despite the higher rents and expensive cost of life in that region, he also noticed that there is a high rate of literate children, which could ensure a good community for his family. 

As he is also an expert \emph{analyst}, he decide to go further in this topic of explaining crime in taxi trips. To achieve this, he seeks to identify additional factors contributing to crime, and analyze whether a standard prediction algorithm faces challenges in that area.

In the crime prediction literature, several hypothetical factors could lead to a good crime prediction. For example, many criminological studies suggest the strong correlation of climate and crime~\cite{anderson2000temperature} and our user is aware about it. He moves to the ``Weather'' layer that contains maximum and minimum temperature, and total precipitation. Using the filters, the user verifies that the highest incidence of crime occurs in the months of February and March, which also coincides with the high temperature in São Paulo during those months.

Many studies perform crime prediction based on the detection of hotspots ~\cite{yang2018crimetelescope}, and he learned about that. Then, he switches to the ``Hotspots'' layer and see that hotspots are present in the majority of corners, which is the granularity of the data, but the difference is the number of occurrences accumulated in the hotspots and using the filter the user is able to realize that right in the indicated location there are 4 hotspots, which is not favorable for their security. \\

Seeking to better understand the data modeling, the user explores the ``Graph'' layer, seeing that each corner is associated to a feature vector holding  information from the underlying layers. He notices that the graph itself provides meaningful information, for instance, the presence of dead ends that can be an indicator of insecurity.  

In the particular case of Sao Paulo, there are some very densely populated settlements or shantytowns, the well-known ``Favelas''. Their residents are usually low-income individuals who face economic challenges, and these communities can be associated with issues such as poverty, crime, and inadequate access to essential services. The user observes that the favelas are mostly on the outskirts of the city, and that there are no favelas near his location. But it also finds a correlation between the ``Favelas'' and ``Trips'' layers, a large number of trips with criminal occurrence depart from or have favelas as their destination.

Finally, the user explores the ``Prediction'' layer and notices that the predictive model fails more in the peripheries, especially in the southwestern part. This area also concentrates a large number of taxi trips with criminal occurrences, and unlike the rest of the city, the number of trips with and without occurrences is balanced. He figures out that the predictive model that generally works well for unbalanced data fails in this balanced region. 

\section{User Study}
The user study was designed to evaluate the effectiveness of layer toggling and dynamic filtering features through a rigorous experimental framework. A total of 23 participants were recruited for the study, each of whom completed both phases of the experiment. 

The study participants were predominantly aged between 18-34 years (15 participants, approximately 68\%) and 35-54 years (7 participants, approximately 32\%), with a gender distribution of 5 women, 17 men, and 1 preferring not to disclose. A majority of the participants (22 individuals) were postgraduate students, primarily pursuing PhDs in fields such as Computer Science, Mathematics, Statistics, and Engineering, indicating a robust foundation in quantitative analysis and research methodologies; among them, 4 identified as developers, which enhances their technical engagement with data visualization tools, while 1 participant served as an assistant professor, offering an academic perspective. Three participants had prior experience in urban data visualization, providing valuable insights and skills relevant to the study, while the remaining participants lacked experience in this area, potentially influencing their interactions with the visualization tools used in the research.

To initiate the study, an instruction phase was conducted, during which a demonstration of the tool was provided. This demonstration included an analysis of a neighborhood, adhering to specific criteria that mirrored the tasks participants would later undertake. Following the demonstration, participants were tasked with completing similar activities while taking notes on a provided form. At the conclusion of this phase, they filled out a questionnaire to capture their feedback and experiences.

\subsection{Study 1: Layer Toggling}

\textbf{Objectives:}
\begin{itemize}
    \item Assess whether enabling and disabling layers reduces visual clutter when compared to displaying all layers concurrently.
    \item Evaluate whether the Button Box device facilitates more efficient layer switching.
\end{itemize}

In the first study, participants were randomly assigned to one out of four groups to evaluate the functionality of toggling between layers, that is, activating or deactivating specific data layers during visual exploration. 
The groups were organized as follows: Group A - without the toggling functionality and button box; Group B - use the mouse to toggle; Group C - use the keyboard to toggle; and Group D - use the Button Box.

The primary goal was to determine whether the ability to enable and disable layers significantly improves the user experience. Participants were tasked with gathering specific data from active layers related to transportation, socioeconomic information, and favelas in the vicinity of Avenida Paulista in São Paulo. They responded to quantitative questions regarding the number of subway stations, average household income, and the presence of favelas within a three-block radius. Their confidence in the answers was measured using a Likert scale from 1 to 5, where 1 indicated low confidence and 5 indicated high confidence. It was anticipated that Group A would require more time to complete the tasks and exhibit signs of discomfort, whereas Groups B, C, and D would perform more efficiently, with Group D expected to achieve the best performance.

\subsubsection{Study 2: Dynamic Filtering}

\textbf{Objectives:}
\begin{itemize}
    \item Determine whether dynamic filters (spatial and temporal) are more efficient than static filters to perform specific tasks.
\end{itemize}

In the second study, all 23 participants engaged in a follow-up experiment aimed at evaluating the effectiveness of dynamic filters in comparison to static filters. The objective of this phase was to ascertain whether dynamic filters (spatial and temporal) offered greater efficiency than the static filters of the original application. Participants were divided into two groups: Group X, which utilized static filters, and Group Y, which employed the proposed dynamic filters, specifically the Dynamic Brush tool and the Temporal Data Dynamic Histogram. 

To identify the block with the highest concentration of crimes, participants utilized the spatial brush filter, which adjusts its size based on the number of crimes. A smaller brush indicates a denser region of crime, facilitating the identification of hotspots. Participants were tasked with analyzing crime data along Avenida Paulista, specifically identifying the block with the highest concentration of crimes and the peak time for these incidents. They were again asked to rate their confidence in their answers on a Likert scale, providing qualitative feedback regarding their experience.

Following the experiments, participants completed a survey to assess their acceptance of the tool. The questionnaire addressed various aspects, including ergonomics, novelty, learning curve, and overall user experience. Additionally, participants who used a device were queried about their cognitive memory related to layer associations and any distractions caused by the auxiliary device.

The distribution of participant groups is summarized in \autoref{tab:user_study}. The details about the two user studies, including the form participants had to answer, are included as supplementary material.

\begin{table}[h]
    \centering
    \caption{User Study Group Distribution: In the first study, participants were assigned to Group A (no device), Group B (mouse), Group C (keyboard), and Group D (Button Box). In the second study, Groups X and Y used static and dynamic filters, respectively.}
    \label{tab:user_study}
    \begin{tabular}{|c|c|c|c|c|}
        \hline
        \textbf{Participants} & \textbf{Group} & \textbf{Functionality} & \textbf{Device} \\ \hline
        \multirow{4}{*}{23} & A (5) & Without functionality & None \\ \cline{2-4} 
        & B (5) & Toggle between layers & Mouse \\ \cline{2-4} 
        & C (5) & Toggle between layers & Keyboard \\ \cline{2-4} 
        & D (8) & Toggle between layers & Button Box \\ \hline
        \multirow{2}{*}{23} & X (10) & Static filters & - \\ \cline{2-4} 
        & Y (13) & Dynamic filters & - \\ \hline
    \end{tabular}
\end{table}

\subsection{Availability of data and materials}
The datasets, modified software, experimental materials, including design documents, videos, forms, and results from the current study, are available at
\url{https://github.com/Kareliavs/Toggler/}. 

\section{Results}
\label{sec:results}

Regarding Study 1, the Kruskal-Wallis test was used to assess the differences in observations for each of the following variables among groups A, B, C, and D: the number of correct answers in the questionnaire (hits), confidence level in responses (confidence), time in seconds to complete (times), and the number of clicks (clicks) required to complete the task. For time measurements, three outliers were identified according to the boxplot criterion — two from group C and one from group D. These outliers were removed, except for one in group C, which, although classified as an outlier, was not considered an extreme value. Regarding the number of clicks, one extreme outlier from group D was also removed. \autoref{tab:KW_results_devices} presents the statistics and p-values from the tests performed.

\begin{table}[h]
\centering
\caption{Kruskal-Wallis (K-W) test statistics and p-values for differences in the following variables between groups A, B, C and D: the number of correct answers in the questionnaire (Hits), confidence level in responses (Confidence), time in seconds (Times), and the number of clicks (Clicks) required to complete the task.}
\begin{tabular}{|c|c|c|}
\hline
\textbf{Variable} & \textbf{K-W Statistics} & \textbf{ P-value}  \\ \hline
Hits                  & 2.0099                    & 0.5703   \\ \hline
Confidence                 & 2.0594                    & 0.5602   \\ \hline
Times            & 11.921                    & 0.0077 \\ \hline
Clicks                & 1.9356                    & 0.5859   \\ \hline
\end{tabular}
\label{tab:KW_results_devices}
\end{table}

\begin{figure}[!ht]
    \centering
    \includegraphics[scale = 0.55]{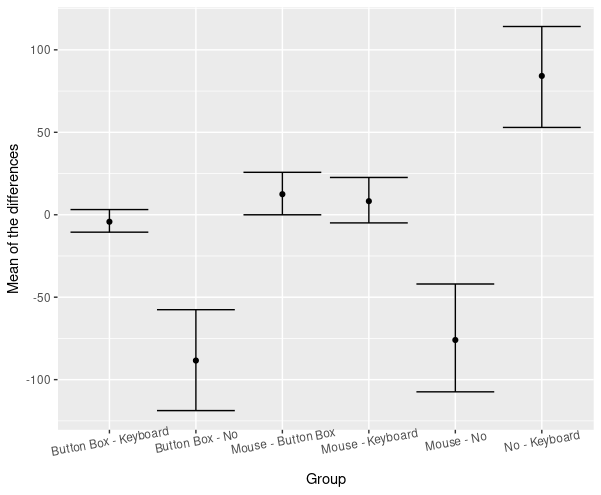}
    \caption{95\% Confidence interval for the difference in means considering all comparisons between groups. For example, the comparison ``Button Box - Keyboard'' represents the difference calculated as the mean of the ``Button Box'' group minus the mean of the ``Keyboard'' group.}
    \label{fig:ci_mean_diff} 
    \end{figure}

The results indicate no significant differences (at the 5\% significance level) in the number of correct answers, confidence levels, or the number of clicks among the groups. However, there is a significant difference in task completion time among participants. The average times (in seconds) for each group were: 107 (A), 31.4 (B), 38 (C), 24.6 (D). To analyze multiple comparisons, a bootstrap study was conducted on the differences in means between groups. For each pair of groups, bootstrap sample means were compared, and the mean of those differences was estimated, along with the 2.5\% and 97.5\% percentiles as confidence intervals for this parameter. The confidence intervals for the difference in means between groups are shown in \autoref{fig:ci_mean_diff}.

It is concluded that task completion times are significantly longer for the group that did not use any device (comparisons ``Button Box - No'', ``Mouse - No'' and ``No - Keyboard'' in \autoref{fig:ci_mean_diff}). For the remaining groups, completion times are considered statistically similar, although their confidence interval limits are close to zero, making the result borderline (comparisons ``Button Box - Keyboard'', ``Mouse - Button Box'' and ``Mouse - Keyboard'' in \autoref{fig:ci_mean_diff}). In this context, since there is no difference in completion time across devices, it can be concluded that the Button Box did not improve the user experience in this regard.

For study 2, the Kruskal-Wallis test was also used to assess differences in observations for each of the previously cited variables between groups X and Y. Outliers and extreme values were only observed for the number of clicks in group Y, and they were removed. \autoref{tab:KW_results_filters} presents the statistics and p-values from the tests performed.

\begin{table}[t]
\centering
\caption{Kruskal-Wallis (K-W) test statistics and p-values for differences in the following variables between groups X and Y: the number of correct answers in the questionnaire (Hits), confidence level in responses (Confidence), time in seconds (Times), and the number of clicks (Clicks) required to complete the task.}
\begin{tabular}{|c|c|c|}
\hline
\textbf{Variable} & \textbf{K-W Statistics} & \textbf{ P-value}  \\ \hline
Hits                  & 0.0491                  & 0.8247   \\ \hline
Confidence                 & 3.6151                    & 0.0573  \\ \hline
Times            & 9.4287                    & 0.0021 \\ \hline
Clicks                & 1.9003                    & 0.1680    \\ \hline
\end{tabular}
\label{tab:KW_results_filters}
\end{table}

\begin{figure}
    \centering
    \includegraphics[width=1\linewidth]{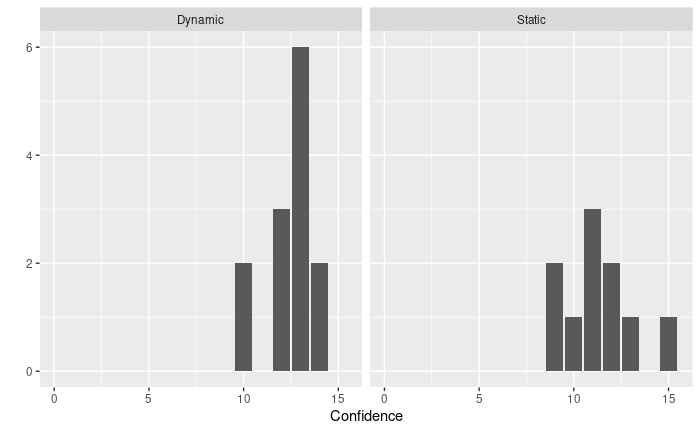}
    \caption{Bar plot of the confidence level demonstrated by the participants in answering the questions about the filters.}
    \label{fig:barplot_confidence}
    \vspace{-0.5cm}
\end{figure}

A borderline difference in confidence levels was observed, with group Y (Dynamic filters) showing higher confidence in their answers, as illustrated in \autoref{fig:barplot_confidence}. No significant difference was found in the number of correct answers or clicks between groups X and Y. However, task completion time was significantly different between the groups at a 5\% significance level, with group Y having the shortest times (in seconds): 131 (X) and 65.7 (Y). This confirms that dynamic filters effectively reduce the time required to complete a task.

Regarding additional questionnaire evaluations, the relationship between the number of 
cases in which a user experience visual information overload and their assigned filter-related group was analyzed. At a 5\% significance level, the hypothesis that the number of overloads does not differs between groups X and Y was not rejected. So, it was concluded that the dynamic filters do not increase or decrease the cases of overcharge compared to the static filters. The study also assessed the relationship between distraction level and ease of use of filters among groups that received different devices. These variables were not found to be significantly different between groups. The results of these tests are presented in the \autoref{tab:KW_add_results}.

\begin{table}
\centering
\caption{Kruskal-Wallis (K-W) test statistics and p-values for differences in the following variables between groups: overload, distraction level and easy of use}
\begin{tabular}{|c|c|c|c|}
\hline
\textbf{Variable} & \textbf{Groups tested} & \textbf{K-W Statistics} & \textbf{P-value} \\ \hline
Overload                       & X and Y                & 0.7591                             & 0.3836           \\ \hline
Distraction level              & A, B, C and D             & 3.5333                             & 0.1709           \\ \hline
Easy of use          & A, B, C and D             & 1.0081                             & 0.6041           \\ \hline
\end{tabular}
\label{tab:KW_add_results}
 \vspace{-0.5cm}
\end{table}

According to the statistical tests presented in this section, the only variable for which the difference between groups was significant was task completion time. This indicates that, although the use (or lack thereof) of devices such as a mouse, keyboard, and button box does not affect users' accuracy in their responses, their confidence levels, or the number of clicks required to complete the task, the use of these devices is associated with a shorter task completion time. However, no difference was observed in completion times among participants who used different devices.

Additionally, a statistical difference was detected in task completion times between participants who were provided with dynamic and static filters; the group using dynamic filters completed the tasks in less time. 

\section{Conclusion}
\label{sec:conclusion}
In this paper, we have presented a novel approach to urban data visualization that addresses the challenges of analyzing multivariate and multimodal datasets, particularly in the context of predictive modeling for crime and mobility in São Paulo. Our key contributions include:
\begin{enumerate}
    \item Layered Visualization Methodology: We proposed a method that organizes urban data into distinct layers, allowing each layer to occupy the entire screen space. This approach enhances spatial resolution and facilitates the perception of both local and global correlations among features.
    \item Dynamic Layer Toggling and Filtering Techniques: Our innovative layer toggling mechanism, combined with dynamic filtering capabilities, enables users to navigate complex datasets efficiently. By allowing users to maintain focus on the screen while comparing features, we enhance the overall user experience.
    \item Comprehensive Evaluation of Predictive Systems: We introduced a framework for visually evaluating predictive models, enabling the identification of areas with successful predictions as well as those with poor performance. This provides valuable insights for further refinement of predictive algorithms.
    \item It was possible to observe through the statistical methods used that the proposed tool for using dynamic filters to improve data visualization indeed enhanced the user experience in terms of the time required to complete a task within the platform. Also, it was concluded that the use of the Button Box device does not reduce task completion time compared to other devices such as a mouse and keyboard. However, these completion times remain equivalents and significantly better than the absence of a device. Although dynamic filters reduced task completion time, there was no evidence that the change increased users' confidence or accuracy in their responses, nor did it decrease the number of clicks required to complete the task. The same was observed for the use of devices, where the inclusion of the Button Box did not appear to enhance confidence, accuracy, or reduce the number of clicks. So, it can be concluded that, although the button box may not be better than other devices considering times of task executions, the user experience was impacted in terms of the time required to complete a task, with the implementation of dynamic filters and the use of any device — whether a keyboard, mouse, or button box — leading to faster execution.

\end{enumerate}
Despite these advancements, there are several potential avenues for future research. First, exploring the integration of real-time data streams could enhance the responsiveness of our visualization system, allowing for more dynamic and timely analyses. Second, expanding our methodology to incorporate additional data types, such as social media interactions or sensor data, could provide a more comprehensive view of urban phenomena. 

Furthermore, investigating user interactions with our visualization tools through user studies could yield insights into usability and effectiveness, leading to further refinements in design. Lastly, applying our approach to other urban contexts could validate its adaptability and effectiveness across different environments and datasets.

In summary, our work contributes to the field of urban data visualization by providing a robust framework for analyzing complex datasets, while also highlighting opportunities for future exploration and innovation.

\section{Acknowledgements}
This work was supported by FAPESP (\#2020/07012-8, \#2022/09091-8). The opinions, hypotheses, conclusions and recommendations expressed in this material are the responsibility of the authors and do not necessarily reflect the views of FAPESP.

\bibliographystyle{unsrt}
\bibliography{references}
\appendix
\section{Experiment Description}
\label{app1}
\addcontentsline{toc}{section}{Appendix A: Experimental Details}



\section*{Supplementary material}

# Instruction Phase

A simplified demonstration of what the tool does will be shown through a story.

Let's assume I'm looking for a residential area that meets the following criteria:

1. Good location
2. Well-connected to transportation networks
3. Household income to determine affordability
4. Safety
5. Additionally, I want to know the best time to avoid going out for safety reasons.

I've been recommended Pinheiros, so I input it in the geocoder and see that it's relatively central, fulfilling the first criterion of a good location.

Next, I activate the transportation layer (I'll demonstrate the three interaction methods: mouse, keyboard, or button box), and I see that there are several bus stops and subway stations. This fulfills the second criterion.

Then, I activate the socioeconomic layer and the tooltip feature, move to the census sector where the marker is located, and find that the average household income is around 7,000 reais.

I activate the crime layer, the crime concentration appears less dense than in the city center. To confirm, I use the brush (demonstrating dynamic spatial filtering), and I notice that the concentration is more scattered in Pinheiros. This fulfills the fourth safety criterion.

For the fifth point, with the crime layer active, I view the distribution of crimes by hour using the temporal histogram (demonstrating the animation feature) and see that the most problematic time is 8 pm.

At this point, I have demonstrated the new features: layer toggling and Visibility Preserving Lenses (dynamic spatial and temporal filters).

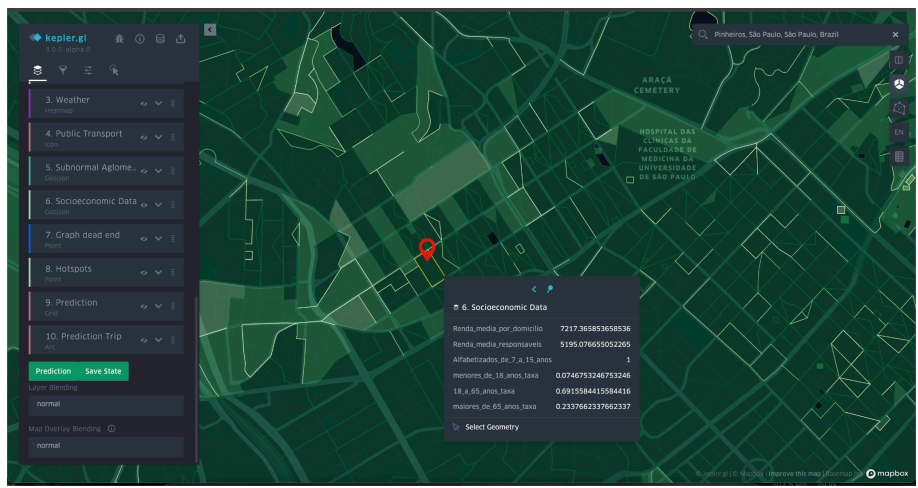

# toggler.gl

# User Experiment Phase

## Layer Toggling

**Objective:**

- Test whether enabling and disabling layers removes visual noise when showing all layers together.
- Test whether the Button Box device facilitates layer switching.

**Methodology:**

- **Participants:** 20 participants divided into 4 groups
- **Functionality:** Toggle between layers
- **Devices:** Without device, with mouse, with keyboard, with Button Box

**Group A:** Without functionality or device: Layers will be overlaid, and participants will attempt to gather the required data.
**Group B:** With functionality to toggle layers using the mouse.
**Group C:** With functionality to toggle layers using the keyboard.
**Group D:** With functionality to toggle layers using the Button Box.

**Expected Results:**

- Group A will take longer to complete the task and show signs of discomfort.
- Groups B, C, and D will perform the task more efficiently.
- Group D will have the best performance.

**Tasks:** The screen will show the active layers for transportation, socioeconomic data, and favelas, with a geolocation marker at Avenida Paulista in São Paulo. Participants will be asked to gather three pieces of data involving interaction with all three layers:

1. How many subway stations are on Avenida Paulista?
2. What is the average household income in the census sector where the marker is located?
3. Are there any favelas within a three-block radius nearby? How many?

For each of the questions, the confidence of the response will be checked.

- How confident are you in your answer? (Likert scale 1-5)
    - 1: Not at all confident, 5: Extremely confident

**Metrics:**

- **Time:** The time from the start of the user's exploration phase will be recorded, using the average time for each group as a reference.
- **Recording:** A video log will be kept for each participant to explain delays or user experience.

- **Likert scale for quantitative questions.**
- **Qualitative questions.**
- **Tracking the number of clicks.**

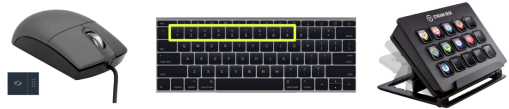

## Visibility Preserving Lenses

**Objective:**

- Test whether dynamic filters (spatial and temporal) are more efficient than the static filters of the original application.

**Methodology:**

- **Participants:** 20 participants divided into 2 groups
- **Functionality:** Use the Dynamic Brush tool and Temporal Data Dynamic Histogram.

**Group X:** Does not use dynamic tools, only static filters.

**Group Y:** Uses dynamic tools.

**Expected Results:**

- Group X will struggle and take longer to complete the task.

**Tasks:** The screen will show the active layer for crimes involving cars and cell phones in 2020, with a geolocation marker at Avenida Paulista in São Paulo. Participants will be asked to complete two tasks:

1. Which block has the highest concentration of crimes along Avenida Paulista? Mention the intersection or reference street.
2. Which is the peak time for crimes on Avenida Paulista?
3. Do you think there is a relationship between crime rates and the proximity of public transportation stations? Please explain your reasoning.

For each of the questions, the confidence of the response will be checked.

- How confident are you in your answer? (Likert scale 1-5)
    - 1: Not at all confident, 5: Extremely confident

**Metrics:**

- **Time:** The time from the start of the user's exploration phase will be recorded, using the average time for each group as a reference.
- **Performance:** Measure if they got the task right

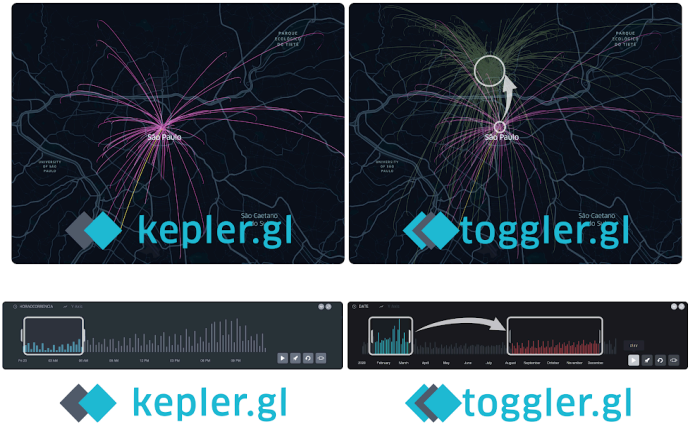

- **Recording:** A video log will be kept for each participant to explain delays or user experience.
- **Tracking the number of clicks.**

## Questionnaire

Following the experiments, participants will complete a survey to assess their level of acceptance.

- [Ergonomics] How much visual overload did you experience with the layers? (Likert scale 1-5)
    - 1: Low overload, 5: High overload
- [Ergonomics] How comfortable did you feel performing the task? (Likert scale 1-5)
    - 1: Uncomfortable, 5: Very comfortable
- [Novelty] How novel do you find the tool? (Likert scale 1-5)
    - 1: Not novel, 5: Very novel
- [Learning curve] How difficult was it to learn how to use the tool? (Likert scale 1-5)
    - 1: Very easy, 5: Very difficult
- [Feedback] Do you have any additional feedback or comments about your experience using the tool? Do you have any improvements or changes to suggest?
- [Feedback] Was the instruction phase clear? Please provide any feedback.

**For those who used a device:**

- [Cognitive memory] Was it easy to associate each layer with its corresponding key/button? (Likert scale 1-5)
    - 1: Very easy, 5: Very difficult
- [Eye fixation] Did the use of the auxiliary device distract you during the task?

toggler.gl

# Summary

In conclusion, there will be 20 users divided into 4 groups for the task of toggling between layers and divided into 2 groups for the task of using the dynamic filters.

| 20 users | | **A** (5) | **B** (5) | **C** (5) | **D** (5) |
|---|---|---|---|---|---|
| Layer Toggling | Functionality | ✘ | ✔ | ✔ | ✔ |
| | Device | ✘ | Mouse | Keyboard | Button Box |
| 20 users | | **X** (10) | | **Y** (10) | |
| Visibility Preserving Lenses | Dynamic filters | ✘ | | ✔ | |

# Experiment implementation

During a week-long trial, the tool was evaluated by 23 participants, including students, researchers, and developers. Each participant engaged in the experiment for 30 minutes, during which they were provided with a response sheet (see Appendix 1) to document their observations while completing the tasks. The execution time for each task, the learning time for the tool (measured per task), and the number of clicks were recorded. Additionally, session logs were captured and stored in a designated folder (see Appendix 2), which also includes handwritten forms in three languages: English, Spanish, and Portuguese (see Appendix 2).

Following the trial, the responses were standardized into a single format by translating all content into English and ensuring consistency, such as standardizing the use of decimal points in the responses. A table of responses was created and is provided in Appendix 3, where correct answers are highlighted in green. Open-ended questions were standardized, with some responses converted to binary format. This standardized response data serves as the input for the subsequent hypothesis testing.

# Appendices

- **Appendix 1**: Response Sheet

- **Appendix 2**: Recorded Sessions Folder and Handwritten Forms in English, Spanish, and Portuguese https://github.com/anonymized.for.reviewing

- **Appendix 3**: Standardized Response Table

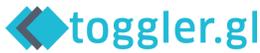

*Indicates a required question.

E-mail *

**1.** Name

**2.** What is your gender? *

*Select only one option.*

○ Female
○ Male
○ Prefer not to say

**3.** What is your age group? *

*Select only one option.*

○ Under 18
○ 18-34
○ 35-54
○ 55 or older

**4.** What is your occupation? *

**Tasks**

**Layer Toggling**

1. How many subway stations are on Avenida Paulista? *

How confident are you in your answer? *
*Select only one option.*

|  | 1 | 2 | 3 | 4 | 5 |  |
|---|---|---|---|---|---|---|
| Not at all confident | ○ | ○ | ○ | ○ | ○ | Extremely confident |

2. What is the average household income in the census sector where the marker is located? *

How confident are you in your answer? *
*Select only one option.*

|  | 1 | 2 | 3 | 4 | 5 |  |
|---|---|---|---|---|---|---|
| Not at all confident | ○ | ○ | ○ | ○ | ○ | Extremely confident |

3. Are there any favelas within a three-block radius nearby? How many? *

How confident are you in your answer? *
*Select only one option.*

|  | 1 | 2 | 3 | 4 | 5 |  |
|---|---|---|---|---|---|---|
| Not at all confident | ○ | ○ | ○ | ○ | ○ | Extremely confident |

**Visibility Preserving Lenses**

1. What block has the highest concentration of crimes along Avenida Paulista? * Mention the intersection or reference street.

How confident are you in your answer? *
*Select only one option.*

|  | 1 | 2 | 3 | 4 | 5 |  |
|---|---|---|---|---|---|---|
| Not at all confident | ○ | ○ | ○ | ○ | ○ | Extremely confident |

2. What is the peak time for crimes on Avenida Paulista? *

How confident are you in your answer? *
*Select only one option.*

|  | 1 | 2 | 3 | 4 | 5 |  |
|---|---|---|---|---|---|---|
| Not at all confident | ○ | ○ | ○ | ○ | ○ | Extremely confident |

3. Do you think there is a relationship between crime rates and the proximity of public transportation stations? Please explain your reasoning. *

How confident are you in your answer? *
*Select only one option.*

|  | 1 | 2 | 3 | 4 | 5 |  |
|---|---|---|---|---|---|---|
| Not at all confident | ○ | ○ | ○ | ○ | ○ | Extremely confident |

**Questionnaire**

1. How much visual overload did you experience with the layers?*

*Select only one option.*

|  | 1 | 2 | 3 | 4 | 5 |  |
|---|---|---|---|---|---|---|
| Low overload | ○ | ○ | ○ | ○ | ○ | High overload |

2. How comfortable did you feel performing the task? *

*Select only one option.*

|  | 1 | 2 | 3 | 4 | 5 |  |
|---|---|---|---|---|---|---|
| Uncomfortable | ○ | ○ | ○ | ○ | ○ | Very comfortable |

3. How novel do you find the tool?*

*Select only one option.*

|  | 1 | 2 | 3 | 4 | 5 |  |
|---|---|---|---|---|---|---|
| Not novel | ○ | ○ | ○ | ○ | ○ | Very novel |

4. How difficult was it to learn how to use the tool? *

*Select only one option.*

|  | 1 | 2 | 3 | 4 | 5 |  |
|---|---|---|---|---|---|---|
| Very easy | ○ | ○ | ○ | ○ | ○ | Very difficult |

5. Do you have any additional feedback or comments about your experience using the tool? Do you have any improvements or changes to suggest?

_____
_____
_____
_____

6. Was the instruction phase clear? Please provide any feedback.

_____
_____

7. **For those who used a device:** Was it easy to associate each layer with its corresponding key/button?

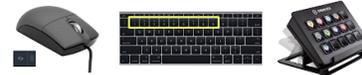

*Select only one option.*

|  | 1 | 2 | 3 | 4 | 5 |  |
|---|---|---|---|---|---|---|
| Very easy | ○ | ○ | ○ | ○ | ○ | Very difficult |

8. **For those who used a device:** Did the use of the auxiliary device distract you during the task?

_____
_____
_____
_____

| Marca temporal | Dirección de correo electrónico | 1. Name | 2. What is your gender? | 3. What is your age group? | 4. What is your occupation? | 1. How many subway stations are on Avenida Paulista? | How confident are you in your answer? | 2. What is the average household income in the census sector where the marker is located? | How confident are you in your answer? | 3. Are there any favelas within a three-block radius nearby? How many? | How confident are you in your answer? | 1. What block has the highest concentration of crimes along Avenida Paulista? Mention the intersection or reference street. | How confident are you in your answer? | 2. What is the peak time for crimes on Avenida Paulista? | How confident are you in your answer? |
|---|---|---|---|---|---|---|---|---|---|---|---|---|---|---|---|
| 15/03/2025 23:17:02 | sajjad.bakrani@icmc.usp.br | Sajjad Bakrani | Male | 18-34 | Researcher | 3 | 4 | 9558 | 4 | 0 | 3 | Uniformly distributed | 3 | 6:30 - 10:30 pm | 4 |
| 15/03/2025 23:28:53 | ximena.pocco@usp.br | Ximena Sofia Pocco Lozada | Female | 18-34 | Researcher | 3 | 5 | 9558 | 4 | 0 | 5 | Passagem Subterrânea Daher | 5 | 6:45 - 8:00 pm | 4 |
| 15/03/2025 23:56:02 | estefanyqch@gmail.com | Estefany | Female | 18-34 | Student | 5 | 4 | 9558 | 5 | 0 | 5 | At the beginning of the avenue | 4 | 6:00 - 7:40 pm | 4 |
| 16/03/2025 17:16:09 | ldwnlp@gmail.com | Ludwin | Male | 35-54 | Student | 2 | 5 | 9558 | 5 | 1 | 5 | Intersection between Paulista and Carlos Sampaio. | 5 | 7:55 - 8:22 pm | 4 |
| 16/03/2025 17:50:18 | manuel.125it@gmail.com | Manuel | Male | 18-34 | Student | 2 | 4 | 14026 | 4 | 0 | 4 | Rua Augusta and Av. Paulista | 4 | 8:00 - 10:00 pm | 5 |
| 17/03/2025 0:13:29 | cristianocc916@gmail.com | Cristian Castillo | Male | 18-34 | Student | 2 | 4 | 9558 | 5 | 0 | 4 | Avenida Paulista and Rua Carlos Sampaio | 4 | 8:50 - 9:30 pm | 4 |
| 17/03/2025 0:26:11 | abrahamsan94@icloud.com | Abraham Rojas Vega | Male | 18-34 | Student | 2 | 4 | 9558 | 4 | 0 | 5 | Rua Leôncion de Carvalho | 3 | 6:22 - 8:10 pm | 4 |
| 17/03/2025 0:37:33 | carlospedraza@usp.br | Carlos | Prefer not to say | 35-54 | Phd Student | 4 | 3 | 14026 | 4 | 0 | 4 | Av. Paulista at the intersection with Rua Carlos Sampaio. | 3 | 7:50 - 8:26 pm | 4 |
| 17/03/2025 0:44:50 | o.laurani@usp.br | Oskar Laurani | Male | 18-34 | Student | 3 | 5 | 9558 | 5 | 0 | 5 | Rua Pamplona - Alameda Rio Claro | 3 | 7:27 - 8:50 pm | 4 |
| 17/03/2025 2:47:15 | andrereliquias.contato@gmail.com | André Reliquias Santo Minas | Male | 18-34 | Developer | 2 | 3 | 9558 | 5 | 0 | 5 | Avenida Paulista x Alameda Joaquim | 3 | 9:00 pm | 4 |
| 17/03/2025 2:56:11 | enriqueteles@usp.br | Enrique Teles | Male | 18-34 | Student | 3 | 5 | 9558 | 5 | 0 | 5 | The block with the intersection on Rua Augusta | 2 | 7:00 - 8:00 pm | 3 |
| 17/03/2025 3:24:48 | joaovneres@usp.br | João Victor Cosme Neres de Sousa | Male | 18-34 | Developer | 3 | 5 | 9558 | 5 | 0 | 5 | At Avenida Paulista with Rua Pamplona. | 4 | 9:00 pm | 5 |
| 17/03/2025 3:35:40 | israel.huaman@ucsp.edu.pe | Israel | Male | 18-34 | Researcher and developer | 3 | 5 | 9558 | 5 | 0 | 5 | From the middle to the south. Cell phone theft on Rua Carlos Sampaio, João Manuel, and Rua Augusta. | 5 | 7:45 - 8:45 pm | 5 |
| 18/03/2025 0:08:13 | nike.codeux@gmail.com | Wilbur Naike Chiuyari Veramendi | Male | 35-54 | Phd Student | 3 | 5 | 9558 | 4 | 0 | 4 | South East | 4 | 6:20 - 9:32 pm | 4 |
| 18/03/2025 0:22:02 | viniciusrpb@unb.br | Vinicius R. P. Borges | Male | 35-54 | Assistant Professor | 2 | 5 | 9558 | 5 | 0 | 5 | Passagem Subterrânea Daher Elias Cutait | 4 | 7:45 - 8:06 pm | 5 |
| 19/03/2025 1:36:59 | gustavoosb@gmail.com | Gustavo Barros | Male | 18-34 | Developer | 2 | 4 | 9558 | 5 | 0 | 4 | Rua Frei Caneca, Rua Carlos Comenale | 4 | 7:45 - 8:17 pm | 4 |
| 19/03/2025 1:45:34 | brepejon@gmail.com | Breno Pejon | Male | 18-34 | Developer | 3 | 5 | 9558 | 5 | 0 | 5 | Rua Paulista with Rua Augusta | 5 | 6:30 - 7:30 pm | 5 |
| 19/03/2025 23:44:29 | priscylla.silva@usp.br | Priscylla Silva | Female | 18-34 | Phd Student | 3 | 5 | 9558 | 5 | 0 | 5 | Intersection of Av. Paulista and Av. Brigadeiro Luís Antônio | 4 | 7:00 - 9:00 pm | 4 |
| 20/03/2025 20:48:24 | marcusvlc@usp.br | Márcus V. L. Costa | Male | 18-34 | Student | 5 | 4 | 9558 | 5 | 0 | 5 | Rua Augusta | 4 | 8:00 - 9:30 pm | 5 |
| 20/03/2025 20:53:22 | samuelmarques@usp.br | Samuel de França Marque | Male | 18-34 | Postdoctoral researcher | 3 | 5 | 9558 | 5 | 0 | 5 | Near the intersection with Rua Augusta. | 4 | Approximately at 8:00 PM. | 3 |
| 21/03/2025 22:43:21 | mariavpdo@gmail.com | Maria Vithória Oliivieira | Female | 18-34 | Student | 2 | 5 | 9558 | 5 | 0 | 5 | Rua Pamplona | 4 | 6:58 - 9:15 pm | 4 |
| 21/03/2025 23:26:09 | luizarodriguescardoso@usp.br | Luiza Rodrigues | Female | 18-34 | Student | 3 | 4 | 1426 | 4 | 0 | 4 | The concentration of crimes appears to be uniform. | 4 | 5:15 - 12 pm | 4 |
| 22/03/2025 14:27:41 | kevin.jv@usp.br | Jonathan Kevin | Male | 18-34 | Student | 4 | 4 | 9558 | 5 | 1 | 4 | All along Paulista Avenue. | 4 | 6:00 - 9:00 pm | 4 |

| 3. Standardized | 3. Do you think there is a relationship between crime rates and the proximity of public transportation stations? Please explain your reasoning. | How confident are you in your answer? | 1. How much visual overload did you experience with the layers? | 2. How comfortable did you feel performing the task? | 3. How novel do you find the tool? | 4. How difficult was it to learn how to use the tool? | 5. Do you have any additional feedback or comments about your experience using the tool? Do you have any improvements or changes to suggest? | 6. Standardized | 6. Was the instruction phase clear? Please provide any feedback. | 7. For those who used a device: Was it easy to associate each layer with its corresponding key/button? | 8. Standardized | 8. For those who used a device: Did the use of the auxiliary device distract you during the task? |
|---|---|---|---|---|---|---|---|---|---|---|---|---|
| Yes | It looks like that | 2 | 1 | 3 | 5 | 3 | I didn't get it how the brush is helpful for the system | Yes | Yes, it was | 2 | No | No distraction |
| Yes | No with all the stations. I saw more crimes in the peak close to the station located in the middle. I guess it's the most frequented station | 4 | 2 | 5 | 5 | 3 |  | Yes | Yes, it was 10/10 | 1 | No | I used gato, no it didn't distract me |
| Yes | Yes, due to what happened with the train or bus stations, they have a higher concentration of people, which is reflected in vandalism or theft. | 4 | 3 | 4 | 5 | 3 | Regarding zooming in on the image or map, a great help could be knowing where an avenue begins and/or ends. | Yes | Yes, it was | 2 | A bit | Regarding zooming in on the image or map, a great help could be knowing where an avenue begins and/or ends. |
| Yes | Yes | 4 | 5 | 5 | 4 | 1 | Maybe it would be good to add a layer of regions categorized as very safe, progressively safe, relatively unsafe, and unsafe. Place images at stations, favelas, etc. When clicking on each sector, it should show the nearest metro station and a score from 0 to 100, comparing the percentage of thefts in that sector with others. | Yes | Yes, it was understandable. |  |  |  |
| Yes | Yes, because public transport points are strategic for thieves to hide and escape. | 4 | 3 | 4 | 4 | 3 | Yes, I would like to have that interface on my mobile device. | Yes | Yes, the instructions are clear. I think that by using the interface a few more times, I would quickly become familiar with it. | 1 | A bit | Well, I had to use the keyboard, but I really wanted to use the mouse. I think it was more about habit than distraction, as I'm used to using the mouse and keyboard together. |
| Yes | Yes, because there are several public transport stations near Rua Carlos Sampaio. | 5 | 5 | 4 | 5 | 3 | Public Services Location (Restaurants, hospitals, apartments, etc.) | Yes | Yes, however, I don't think I would have been able to complete the task without instructions. | 2 | No | I used the keyboard; the numbers made it easier to switch layers. |
| No | No, because there are stations with less crimes than other parts of the avenue. | 4 | 2 | 2 | 2 | 3 | Less details on the map (no display individual buildings). Maybe use more brighter colors | Yes | Very clear | 3 | No | Not distracting |
| No | There is no pattern. | 3 | 3 | 4 | 3 | 4 | My eyes struggled to see the tabs; when I hovered over each menu, there should be a zoom. |  | She was brief, she explained the tab activation quickly. |  |  |  |
| Yes | Yes, during peak hours, there is an increase in crime points near the stations. | 3 | 4 | 5 | 5 | 3 | When selecting an area (Av. Paulista, for example), it should be highlighted. | Yes | Yes, the instructions were clear. | 1 | No | No |
| Yes | Yes, criminals can use public transportation to escape the location. | 3 | 2 | 5 | 5 | 2 | I liked the tool given the grouping of various types of data for visualization of the location. | Yes | Yes | 1 | No | No |
| Yes | Yes, there is, the regions near the public transportation have more density of people. The brush tool helped with this conclusion. | 4 | 4 | 4 | 4 | 2 | I believe emojis would help with the layers overload. I find really useful the not static approach with the brush and the crime histogram, but I tested using the static approach. | Yes | The instructions were clear | 1 | No | The keyboard number layer didn't distract me |
| Yes | Yes, analyzing the interface, it can be observed that both on Avenida Paulista and in other locations, there are more crimes near public transportation. This likely happens because criminals know where there will be more people. | 4 | 5 | 5 | 5 | 2 | No, I found it clear and intuitive. I would just add an option to speed up or slow down the moment of passing the crime hours. | Yes | Yes | 1 | Yes | Yes |
| Yes | Yes, there is a relation. On Rua Augusta, there is a metro station parallel to the concentration of crimes, and on Sampaio, there is a congested area with crimes. | 5 | 3 | 5 | 3 | 1 | The time filter takes up a lot of space; I would add the option to resize it. Also, a legend that can be moved on the screen and resized, as the location on the right is not natural for everyone. | Yes | Yes, it was clear, but a video or text would make it easier. |  |  |  |
| Yes | I think there is a relationship because, since it is a main avenue, many public transportation users, after visiting various services (stores, etc.) near Avenida Paulista, return home using these mentioned means of transportation. | 4 | 3 | 4 | 5 | 3 | I don't have any feedback; the tool is very intuitive regarding the requested tasks. | Yes | Yes, the instructions were clear enough. | 1 | No | I had no distractions in this regard; I was able to operate the device without difficulty. |
| No | In case of Av. Paulista, I believe that is not a relationship when compared to other less crowded streets and avenues. I visualize that crowded avenues and streets have thefts across them while those less crowded are more likely to have issues near public transportation stations. | 4 | 3 | 5 | 4 | 1 | Hover of 6. For socioeconomic data, print the plot of the float numbers with precision limited to two decimal points. Replace variable name style (ex. "menored_de_18_anos_taxa") with "taxa de menores de 18 anos", i.e., in natural language. Replace the squares in the menu with the buttons (icons) of Stream Deck. | Yes | They were excellent | 1 | No | I believe no |
| Yes | Yes, using the brush tool, I believe the radius slightly decreases around the stations. | 2 | 2 | 4 | 5 | 2 | I found it very interesting. The only improvement I noticed was that sometimes it was difficult to select a range in the time filtering. | Yes | Yes | 1 | No | No |
| Yes | It seems that there is a crime hotspot around metro stations, as there is a higher density of crimes in these areas compared to bus stop areas or areas without public transportation. | 4 | 3 | 5 | 5 | 3 | I really liked the tool. I would like it to have a filtering option for crimes by street or neighborhood. | Yes | Yes, the instructions were clear and well explained. | 3 | No | My difficulty was memorizing which keyboard buttons corresponded to which layer of information. However, this information was accessible on the screen to check and know which button to press. |
| Yes | Outside of peak hours, yes. During peak hours, there doesn't seem to be a relationship. | 4 | 3 | 4 | 4 | 3 | The trip layer is a bit confusing to interpret and visualize. | Yes | Yes. Suggestion: It would be good for the participant to explore with the tool beforehand. | 1 | No | No |
| Yes | Yes, due to the higher concentration of people between train stations and bus stops. | 5 | 1 | 5 | 5 | 4 | There could be a button for the light mode of the application. | Yes | All phases | 1 | No | No, Stream Deck was the best device for me, and the mouse is useful for zooming between the layers. |
| Yes | Yes, there is a high concentration of occurrences in regions with many stations. | 4 | 2 | 5 | 4 | 2 | Include the temporal variation of crimes by region (buffer). | Yes | Yes |  |  |  |
| No | No, I think the distribution of crimes on Av. Paulista is uniform | 4 | 3 | 5 | 5 | 3 | To create the same tool for other states in Brazil. | Yes | Yes |  |  |  |
| Yes | Yes, because during these times there is more movement of people on the avenue, who travel via public and private transportation. This makes them more susceptible to crimes in that area. | 4 | 1 | 4 | 5 | 3 | No | Yes | Yes | 2 | No | No |
| Yes | There is a high number of cell phone thefts all along Paulista Avenue, especially around bus stops. | 5 | 1 | 5 | 5 | 4 | Lighter colors. | Yes | Perfectly clear. | 4 | No | No |

| Device | Filter | # clics 1 | # clics 2 | Learning Time 1 | Time 1 | Learning Time 2 | Time 2 | Learning Time 3 | Time 3 | Learning Time 4 | Time 4 | Learning Time 5 | Time 5 | Learning Time 6 | Time 6 |
|---|---|---|---|---|---|---|---|---|---|---|---|---|---|---|---|
| Mouse | Static | 8 | 40 | 0:30 | 0:10 | 0:09 | 0:06 | 0:42 | 0:08 | 0:40 | 0:30 | 0:44 | 0:30 | 1:00 | 0:45 |
| Button Box | Dynamic | 6 | 33 | 1:23 | 0:15 | 1:25 | 0:05 | 0:07 | 0:07 | 0:10 | 0:05 | 1:00 | 0:32 | 1:00 | 0:20 |
| Mouse | Dynamic | 19 | 50 | 1:25 | 0:35 | 0:18 | 0:10 | 0:11 | 0:04 | 0:36 | 0:08 | 0:17 | 0:10 | 0:20 | 0:10 |
| No | Static | 16 | 50 | 0:00 | 0:20 | 0:00 | 0:13 | 0:00 | 0:35 | 1:00 | 0:34 | 0:17 | 0:16 | 0:50 | 0:25 |
| Keyboard | Dynamic | 7 | 28 | 0:11 | 0:06 | 0:22 | 0:20 | 0:07 | 0:03 | 1:15 | 0:20 | 0:44 | 0:16 | 0:07 | 0:03 |
| Keyboard | Dynamic | 4 | 23 | 0:08 | 0:02 | 0:08 | 0:08 | 0:12 | 0:03 | 0:10 | 0:05 | 0:24 | 0:05 | 0:45 | 0:15 |
| Mouse | Static | 31 | 41 | 0:50 | 0:05 | 0:27 | 0:03 | 0:09 | 0:05 | 0:26 | 1:30 | 0:34 | 0:46 | 0:13 | 0:23 |
| No | Static | 11 | 37 | 0:00 | 0:25 | 0:00 | 0:30 | 0:00 | 1:00 | 0:30 | 1:03 | 0:20 | 0:10 | 0:30 | 0:50 |
| Button Box | Static | 4 | 34 | 0:08 | 0:09 | 0:52 | 0:02 | 0:04 | 0:06 | 0:10 | 0:22 | 0:07 | 0:20 | 0:10 | 0:20 |
| Button Box | Dynamic | 7 | 24 | 0:13 | 0:05 | 0:02 | 0:05 | 0:03 | 0:03 | 0:40 | 0:10 | 0:32 | 0:08 | 0:15 | 0:05 |
| Keyboard | Static | 9 | 24 | 0:05 | 0:14 | 0:22 | 0:06 | 0:15 | 0:05 | 0:36 | 0:54 | 0:37 | 1:13 | 0:34 | 0:26 |
| Mouse | Dynamic | 10 | 14 | 0:03 | 0:18 | 0:20 | 0:04 | 0:22 | 0:02 | 1:10 | 0:25 | 0:26 | 0:05 | 0:15 | 0:25 |
| No | Static | 22 | 30 | 0:00 | 0:43 | 0:00 | 0:45 | 0:00 | 1:00 | 0:47 | 0:58 | 0:22 | 0:40 | 0:17 | 1:00 |
| Mouse | Dynamic | 19 | 50 | 0:15 | 0:05 | 0:07 | 0:30 | 0:26 | 0:12 | 2:40 | 0:40 | 1:26 | 0:24 | 1:18 | 0:37 |
| Button Box | Dynamic | 107 | 131 | 1:40 | 0:50 | 2:50 | 0:10 | 0:41 | 0:04 | 2:07 | 0:25 | 0:53 | 0:15 | 0:17 | 0:37 |
| Button Box | Dynamic | 20 | 50 | 0:09 | 0:10 | 0:08 | 0:06 | 0:27 | 0:09 | 1:27 | 0:08 | 1:07 | 0:30 | 0:44 | 0:50 |
| Keyboard | Dynamic | 22 | 135 | 0:14 | 0:30 | 0:45 | 0:45 | 0:28 | 0:22 | 0:47 | 0:53 | 1:10 | 0:40 | 1:00 | 1:00 |
| Keyboard | Static | 13 | 23 | 1:08 | 0:12 | 0:16 | 0:07 | 0:25 | 0:07 | 0:07 | 0:21 | 0:05 | 0:48 | 0:05 | 1:00 |
| Button Box | Dynamic | 23 | 27 | 0:16 | 0:05 | 0:15 | 0:08 | 0:11 | 0:04 | 0:50 | 1:00 | 0:18 | 0:20 | 0:14 | 0:30 |
| No | Static | 20 | 78 | 0:00 | 1:30 | 0:00 | 0:30 | 0:00 | 0:21 | 0:02 | 1:07 | 0:30 | 0:23 | 0:45 | 1:15 |
| No | Static | 5 | 60 | 0:00 | 0:25 | 0:00 | 0:10 | 0:00 | 0:30 | 0:34 | 1:00 | 0:20 | 0:40 | 0:25 | 1:20 |
| Button Box | Dynamic | 5 | 36 | 0:41 | 0:06 | 0:08 | 0:12 | 0:13 | 0:02 | 0:18 | 0:50 | 1:00 | 0:10 | 0:07 | 0:05 |
| Button Box | Dynamic | 18 | 18 | 0:36 | 0:04 | 0:07 | 0:05 | 0:45 | 0:05 | 0:35 | 0:05 | 0:10 | 0:20 | 0:26 | 0:10 |







\end{document}